\documentclass[preprint,amsmath,amssymb]{revtex4-1}

\usepackage{graphicx}
\usepackage{dcolumn}
\usepackage{bm}
\usepackage{color,graphicx}
\usepackage[utf8]{inputenc}

\begin{document}

\title{Magnetic and geometric effects on the electronic transport of {metallic} nanotubes}

\author{Felipe Serafim$^\dagger$}
\author{Fernando A. N. Santos$^\star$}
\author{Jonas R. F. Lima$^{\dagger \|}$}
\author{S\'ebastien Fumeron$^\ddag$}
\author{Bertrand Berche$^\ddag$}
\author{Fernando Moraes$^\dagger$}
\affiliation{$^\dagger$Departamento de F\'{\i}sica, Universidade Federal Rural de Pernambuco, 52171-900, Recife, PE, Brazil}
\affiliation{$^\star$Departamento de Matem\'atica,  Universidade Federal  de Pernambuco, 50670-901, Recife, PE,
 Brazil}   
\affiliation{$^\ddag$Laboratoire de Physique et Chimie Th\'eoriques, UMR Universit\'e de Lorraine - CNRS 7019, 54506 Vand\oe uvre les Nancy, France }
\affiliation{$^\|$Institute of Nanotechnology, Karlsruhe Institute of Technology, D-76021 Karlsruhe, Germany}

\date{\today}

\begin{abstract}
{The investigation of curved low-dimensional systems is a topic of great research interest. Such investigations include two-dimensional systems with cylindrical symmetry.}
In this work, we present a numerical study of the electronic transport properties of  {metallic} nanotubes deviating from the cylindrical form either by having a bump or a depression, and under the influence of a magnetic field. Under these circumstances, it is found that the  nanotube may be used as an energy high-pass filter for electrons. It is also shown that the device can be used to tune the angular momentum of transmitted electrons. 
\end{abstract}

\maketitle

\section{Introduction}

Low-dimensional carbon-based materials have long fascinated chemists and physicists alike. From fullerene \cite{KROTO}, which was the object of the year 1995 Chemistry Nobel prize, to  polyacetylene \cite{C39770000578,PhysRevLett.39.1098}  (2000 Chemistry Nobel prize),  to graphene \cite{Novoselov} (2010 Physics Nobel prize), these zero-, one- and two-dimensional forms of carbon certainly deserve all the attention they have had due to their unique properties and numerous technological applications. Carbon nanotubes (CNTs) \cite{Iijima1991}, even though are not directly attached to a Nobel award, are of no less importance for similar reasons. {Since the first proposals of CNTs \cite{OBERLIN1976335}, there is a great interest in the investigation of quantum systems with cylindrical symmetry. The list of nanotubes synthesized and theoretically proposed of different materials increases every year and they show a variety of electronic properties \cite{Tenne1992,Feldman222,PhysRevB.49.5081,Chopra966,doi:10.1021/jacs.7b01652,CRADWICK1972,C0CP00438C,C4RA08840A,Monet2018,Zhang2372}. For instance, WS$_2$ nanotubes are semiconducting, but with doping, it is possible to turn them metallic or superconducting \cite{Qin2017,doi:10.1021/acs.nanolett.8b02647}. We can also cite boron nitride nanotubes, which are insulators with a constant band-gap independent of their radius and helicity \cite{Blase_1994}. On the other hand, depending on the radius and helicity, CNTs can be metallic or semiconductors \cite{saito1998physical}.}

{The investigation of two-dimensional electron gases with cylindrical symmetry (C2DEGs) also attracted great deal of attention \cite{PhysRevB.78.115326,PhysRevB.82.205305,VOROBEV2004171,FILGUEIRAS20152110,Geiler1999,Friedl_2008}. For instance, one of the first theoretical calculations of the quantum Hall effect in two-dimentional electron gases (2DEGs) considered a system with cylindrical symmetry \cite{PhysRevB.23.5632}. Such C2DEG was already obtained experimentally by different methods \cite{PRINZ2000828,Westwater,Schmidt2001,LORKE2003347,M_rtensson_2003,doi:10.1063/1.2433040,PhysRevB.75.205309,https://doi.org/10.1002/smll.200701091}, with a radius that can varies from tens of nanometers up to several microns. Even though the typical diameter of single walled CNT is in the few nanometers range, the outer nanotube of wide multiwalled CNTs can have a radius of the same order of magnitude of those obtained for C2DEGs.}

This work deals with electronic transport in nanotubes. We investigate how one can use a combination of geometry and magnetic field to fine-tune electronic transmission across the length of the tube.   Previously, we investigated \cite{santos} how geometry alone can influence the electronic transport on nanotubes. In that work, we found that simple deformations, like the ones shown in Figs. \ref{1bump} and \ref{1dep} below, can have a deep impact in the transmittance profile. Moreover, periodic deformations clearly induce a gap in the energy spectrum, suggesting the use of deformed nanotubes as electronic filters. This geometric tuning of electronic properties of 2D materials has been explored in different contexts by different authors,  \cite{cortijo2007electronic,PhysRevB.79.235407,ortix2011curvature,liang2016coherent}, including ourselves \cite{SERAFIM2019139}, to cite a few.  {With the inclusion of the magnetic field and the angular momentum states, the results presented here extend and complement the ones reported in Ref. \cite{santos} and offer a possibility of experimental verification of the theory proposed in \cite{Ferrari}. The generic features of the results, as the shape of the transmittance and geometric potential curves,  apply {directly to C2DEGs, but also} to metallic nanotubes in general (carbon or not), even though we used CNT parameters. } {Furthermore, even though the electronic properties of semimetallic CNTs are usually obtained from the 2D Dirac equation, metallic CNTs own they high conductivity  to ballistic (nearly free) electron transport, which justifies our use of the Schr\"odinger equation.}

The effect of an applied magnetic field on the electronic transport  of nanotubes has long been studied  (see for instance \cite{ando1997quantum,roche2000aharonov,rosales2007magnetic}). The combination of curvature and electromagnetic field on the properties of charged particles bound to a surface was studied by Ferrari and Cuoghi \cite{Ferrari} who derived the corresponding Schr\"odinger equation. They extended  the da Costa approach \cite{dacosta},  which reveals an effective geometric potential due to curvature, to include an externally applied electromagnetic field. Their approach has been applied to the study of charged particles on a variety of surfaces under applied electromagnetic fields  \cite{ferrari2008cylindrical,silva2015quantum, schmidt2019exact,schmidt2020exact}. Here, we   study the effects of a magnetic field in the electronic transport of deformed nanotubes. We solve numerically the Ferrari-Cuoghi  Schr\"odinger equation with open boundary conditions (see \cite{santos} or \cite{Marchi} for details of the methodology {respectively applied to corrugations or single constrictions}) in order to find the transmittance as a function of injection energy and orbital angular momentum. Then, we analyze the data noting that the deformations allied to the magnetic field make the nanotube a selective tool for the control of the outgoing electron energy and angular momentum.

\section{Derivation of the Ferrari-Cuoghi  Schr\"odinger equation for a cylindrical surface}

In this section, we follow the steps of Ref. \cite{Ferrari} and find a Schr\"odinger equation for a spinless charged particle bound to a cylindrical surface in the presence of a magnetic field.
Let us first write the Schr\"odinger equation with the covariant derivatives coupled with the electromagnetic four-potential. The spatial gauge covariant derivative  is defined as
\begin{equation} \label{Derivadacovariantedegauge}
D_j  = \nabla_j -\frac{iQ}{\hbar}A_j,
\end{equation}  
where $Q$ is the charge of the particle and $A_j$ is the $j$ component of the vector potential, with $j=1,2,3$. The covariant derivative $\nabla_j$ acting on a vector field $v^i$ is given by
\begin{equation}\label{Derivadacovariante}
\nabla_j  v^i= \partial_j v^i+ \Gamma_{jk}^{i}v^k,
\end{equation}
where $\Gamma_{jk}^{i}$ are the Christoffel symbols and $\partial_j$  is the derivative with respect to the spatial variables  $q_j$. The temporal gauge covariant derivative is given by 
\begin{equation}
D_0=\partial_t-\frac{iQ}{\hbar}A_0,
\end{equation}
where $A_0=-V$, with $V$ being the scalar potential (note that we use the signature $(-,+,+,+)$ for the Minkowski metric). So, the Schr\"odinger equation becomes
\begin{eqnarray} \label{eqS2} 
i\hbar D_0 \psi &=& - \frac{\hbar^2}{2m}G^{ij}D_i D_j \psi,
\end{eqnarray}
where $G^{ij}$ is the inverse of the metric tensor. The gauge invariance can be demonstrated with respect to the following gauge transformations:
\begin{equation} \label{transformacaodegauge}
A_j\rightarrow A^{'}_j=A_j +\partial_j \gamma; \hspace{3mm}  A_0\rightarrow A^{'}_0=A_0 +\partial_t \gamma;   \hspace{3mm} \psi \rightarrow \psi^{'}= \psi e^{iQ\gamma/ \hbar},
\end{equation}
where $\gamma$ is a scalar function of space and time.

Substituting Eq. (\ref{Derivadacovariantedegauge}) into Eq. (\ref{eqS2}) we get
\begin{equation}
i\hbar D_0 \psi = - \frac{\hbar^2}{2m}G^{ij} \left[\left( \nabla_i-\frac{iQ}{\hbar}A_i \right) \left( \nabla_j-\frac{iQ}{\hbar}A_j \right) \right] \psi.
\end{equation}
Applying each term in parentheses on $\psi$ leads us to
\begin{eqnarray}
i\hbar D_0 \psi = \frac{1}{2m}[ - \hbar^2 \nabla_i(G^{ij}\nabla_j \psi)&+&iQ\hbar G^{ij}(\nabla_i A_j)\psi + iQ\hbar G^{ij} A_j (\nabla_i\psi) \nonumber \\
 &+& iQ\hbar  G^{ij}A_i(\nabla_j \psi)+Q^2G^{ij} A_i A_j\psi ].
\end{eqnarray}
Substituting the covariant derivative defined in Eq. (\ref{Derivadacovariante}) and with some algebra we obtain that
\begin{equation} \label{eqSc}
i\hbar D_0 \psi =  \frac{1}{2m}\left[ - \frac{\hbar^2}{\sqrt{G}}\partial_{i}(\sqrt{G}G^{ij}\partial_j \psi)+\frac{iQ\hbar}{\sqrt{G}}\partial_i(\sqrt{G}G^{ij}A_j)\psi +2iQ\hbar G^{ij}A_j\partial_i\psi +Q^2 G^{ij}A_i A_j \psi \right],
\end{equation}
where $G = \textrm{det} (G^{ij})$. This is the covariant Schr\"odinger equation for any three-dimensional curvilinear coordinate system when an electric field and a magnetic field are applied. It is worth noting that Eq. (\ref{eqSc}) with $\vec{A} = (A_1, A_2 , A_3)$ is valid for any gauge choice.

\subsection{Application of the da Costa procedure }

Let us now apply in Eq. (\ref{eqSc}) the procedure first proposed by da Costa \cite{dacosta}. Consider a surface $ S $ of parametric equations $ \vec{r} = \vec{r} (q_1, q_2) $, where $ \vec{r} $ is the position vector of an arbitrary point on the surface. The three-dimensional space in the neighborhood of $ S $ can be parametrized as
\begin{equation}
\vec{R}(q_1,q_2,q_3)= \vec{r}(q_1,q_2)+q_3\vec{n}(q_1,q_2),
\end{equation}
where $\vec{n}$ is a vector normal to the surface. The relation between the three-dimensional metric tensor $ G_{ij} $ and the two-dimensional one $ g_{ab}$ is given by
\begin{equation}
G_{ab}= g_{ab}+[\alpha g+ (\alpha g)^T]_{ab}q_3 + (\alpha g \alpha^T)_{ab}q^{2}_{3}; \hspace{3mm} G_{a3}=G_{3a}=0; \hspace{3mm} G_{33}=1,
\end{equation}
where 
\begin{equation}
g_{ab}=\partial_a\vec{r} \partial_{b}\vec{r},
\end{equation}
and $\alpha_{ab}$ is the Weingarten curvature matrix of the surface.

We now introduce a potential $ V_{\lambda} (q_3) $ that confines the particle to the surface $S$, where $ \lambda $ is the parameter that measures the strength of the confinement. This means that in the limit $\lambda \rightarrow \infty$ the wave function will be nonzero only in the vicinity of $q_3 =0$. The goal of this procedure is to obtain a wave function that depends only on $ q_1 $ and $ q_2 $, which are the coordinates on the surface. With this purpose, we introduce a new factorized wave function
\begin{equation}   
\chi(q_1,q_2,q_3)=\chi_{s}(q_1,q_2)\chi_{n}(q_3),
\end{equation}
where the index $s$ indicates components tangent to the surface, and $ n $ the normal one. The transformation $\psi \rightarrow \chi$ is given by
\begin{equation}   
\psi(q_1,q_2,q_3)=[1+\textrm{Tr}(\alpha)q_3+\textrm{det}(\alpha)q_3^2]^{-1/2}\chi(q_1,q_2,q_3).
\end{equation}
So, taking into account the effects of the potential $V_{\lambda} (q_3)$, we can apply the limit $ q_3 \rightarrow 0 $ in the Schr\"odinger equation. With all these considerations, we can obtain that
\begin{eqnarray}\label{eqS3}
i \hbar D_0 \chi  = \frac{1}{2m} \left[ -\frac{\hbar^2}{\sqrt{g}}\partial_{a}(\sqrt{g}g^{ab}\partial_{b}\chi)+\frac{iQ\hbar}{\sqrt{g}}\partial_{a}(\sqrt{g}g^{ab}A_{b}\chi)+2iQ\hbar g^{ab}A_a \partial_{b} \chi \right.  \nonumber \\ + Q^2(g^{ab}A_a A_b+(A_3)^2)\chi -\hbar^2(\partial_3)^2\chi+iQ\hbar(\partial_3 A_3)\chi\nonumber \\ +\left. 2iQ\hbar A_3(\partial_3 \chi)-\hbar^{2}\left(\left[\frac{1}{2}\textrm{Tr}(\alpha) \right]^2-\textrm{det}(\alpha) \right)\chi \right] +V_{\lambda}(q_3)\chi.
\end{eqnarray}
Note that in Eq. (\ref{eqS3}), there is no term mixing $ A_j $ ($ j = 1,2,3 $) and the curvature matrix $ \alpha_{ab} $. This is an evidence that the magnetic field does not couple with the curvature of the surface, regardless of the shape of the surface, the magnetic field, and the gauge choice. We can see in Eq. (\ref{eqS3}) the appearance of the known geometric potential \cite{dacosta}
\begin{equation}
V_{geo}(q_1,q_2)=-\frac{\hbar^2}{2m}\left(\left[ \frac{1}{2}\textrm{Tr}(\alpha) \right]^2-\textrm{det}(\alpha)\right), \label{Vgeo}
\end{equation} 
where the first term is the square of the mean curvature and the second is the Gaussian curvature. {The term between parentheses can be written in terms of the principal curvatures $\kappa_1$ and $\kappa_2$ as  $(\kappa_1-\kappa_2)^2$, which is always positive \cite{dacosta}.}

Defining a new metric tensor  $\tilde{G}$ as
\begin{equation}
\tilde{G}=\left(\begin{array}{ccc} 
g_{11} & g_{12} & 0\\
g_{21} & g_{22} & 0 \\
0 & 0 & 1
\end{array} \right),
\end{equation}
we can rewrite Eq. (\ref{eqS3}) in the compact form
\begin{equation}
i \hbar D_0 \chi = \frac{1}{2m} \tilde{G}^{ij}\tilde{D}_i \tilde{D}_j \chi +V_{geo}\chi + V_{\lambda}(q_3)\chi.
\end{equation}
We can see that Eq. (\ref{eqS3}) can not be separated into two equations, one that would depend only on the tangent coordinates $(q_1, q_2)$ and other that would depend on the normal coordinate $q_3$, because the term $A_3(q_1,q_2,0)\partial_3 \chi$ couples the dynamics along $q_3$ with that along $(q_1, q_2)$. However, we can choose a gauge that cancels $A_3$, thus eliminating this term. From Eq. (\ref{transformacaodegauge}), we can see that the most suitable choice for $ \gamma $ is
\begin{equation} \label{gamma}
\gamma(q_1,q_2,q_3)=-\int^{q_3}_{0} A_3(q_1,q_2,z)dz,
\end{equation}
which makes $A^{'}_{3}=0$, $\partial_{3} A^{'}_{3}=0$ when the limit  $q_3\rightarrow 0$ is taken, while the components $A_1$ and $A_2$ remain unchanged. Thus, we can now separate Eq. (\ref{eqS3}) into the equations
\begin{equation} \label{esq3}
i\hbar \partial_{t}\chi_{n}=-\frac{\hbar^2}{2m}(\partial_{3})^2\chi_{n}+V_{\lambda}(q_3)\chi_{n},
\end{equation}
\begin{eqnarray} \label{esq1q2}
i\hbar \partial_{t}\chi_{s}= \frac{1}{2m} \bigg[- \frac{\hbar^2}{\sqrt{g}}\partial_a (\sqrt{g}g^{ab}\partial_b \chi_{s}) &+& \frac{iQ\hbar}{\sqrt{g}}\partial_a (\sqrt{g}g^{ab}A_b)\chi_s + 2iQ\hbar g^{ab}A_a\partial_b \chi_s + \nonumber \\ &+&Q^2 g^{ab} A_a A_b \chi_s \bigg] + V_{geo} \chi_s + QV\chi_s.
\end{eqnarray}
Eq. (\ref{esq3}) is the one-dimensional Schr\"odinger equation for a particle bound by the potential $ V_{\lambda}(q_3)$. Eq. (\ref{esq1q2}) is the Schr\"odinger equation describing the dynamics of a particle with mass $m$ and charge $Q$ attached to the surface under the effects of electromagnetic fields. These equations demonstrate that the uncoupling between the tangential and normal surface dynamics is only possible with an appropriate gauge choice.

In the next section we will construct Eq. (\ref{esq1q2}) for a surface with cylindrical symmetry, since we are interested in nanotubes. It is important to mention that in Ref.\cite{Ferrari}, the Eq. (\ref{esq1q2}) was already written in spherical and toroidal coordinates.

\subsection{Application to the cylindrical surface}

\begin{figure}[!htb]
\centering
\includegraphics[scale=0.4]{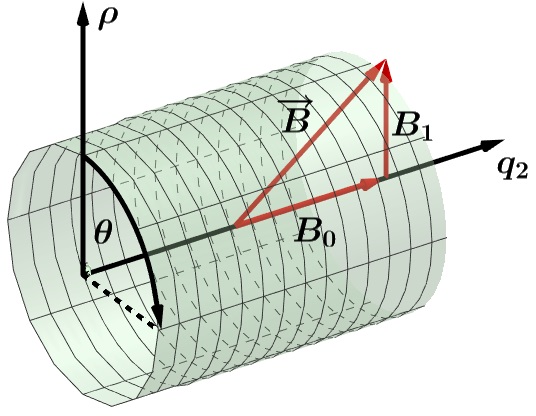}
\caption{Decomposition of the magnetic field.}
\label{TuboMagnetico}
\end{figure}

For a given coordinate system $(\theta, q_2, \rho)$, where our coordinate $q_2$ is the coordinate along the axis of revolution, and $\theta$ is the angular component, a uniform magnetic field $\overrightarrow{B}$ applied to a cylinder of radius $\rho$ can always be decomposed into a component $B_0$ in the direction of $q_2$, parallel to the axis of the cylinder, and a perpendicular component $B_1$. This latter component defines the direction from which the azimuthal angle $\theta$ is defined. It is along the $\rho$-axis at $\theta = 0$, as shown in  Fig. \ref{TuboMagnetico}. This way, the magnetic field is given by $\vec{B}=(B_{\theta},B_{q_{2}},B_{\rho})=(-B_1\sin\theta, B_0, B_1 \cos\theta)$.

Using $\vec{B}=\vec{\nabla} \times \vec{A}$, it is possible to obtain a vector potential 
$\vec A'=\frac 12\vec B\times\vec r$
\begin{equation} \label{potencialvetor}
\vec{A}= (A_{\theta},A_{q_{2}},A_{\rho})=\frac 12 \left(\rho B_0 - B_1 q_2\cos\theta, \rho B_1 \sin\theta , - B_1 q_2\sin\theta \right),
\end{equation}
but in order to simplify the following equations, it is more convenient to proceed to a gauge change $\vec A=\vec A'+\vec\nabla f(\theta,q_2,\rho)$ with $f(\theta,q_2,\rho)=\frac 12 B_1\rho q_2\sin\theta$ which kills one of the components,
\begin{equation}
    \vec{A}= (A_{\theta},A_{q_{2}},A_{\rho})= \left(\frac{1}{2}\rho B_0, \rho B_1 \sin\theta ,0\right).
\end{equation}
We can see clearly that $\vec{\nabla} \times \vec{A}$ gives the correct magnetic field. 

The metric tensor $ g_{ ab}$ and its inverse $ g_{ab}$ are given by
\begin{equation}
g_{ab}=\left( \begin{array}{cc}
\rho(q_2)^2 & 0 \\
0 & 1+\rho^{'}(q_2)^2
\end{array} \right)
\end{equation}
and
\begin{equation} \label{metricainversa}
g^{ab}=\left( \begin{array}{cc}
(\rho(q_2)^2)^{-1} & 0 \\
0 & ( 1+\rho^{'}(q_2)^2)^{-1}
\end{array} \right),
\end{equation}
where $\rho^{'} = d \rho / d q_2$. We also have that
\begin{equation} \label{raizdeg}
\sqrt{g}= \sqrt{\textrm{det}g} = \rho(q_2)\sqrt{1+\rho^{'}(q_2)^2}.
\end{equation}
We now have the indexes $a=q_2$ and $b=\theta$. So, $\chi_s$  becomes a function of $q_2$ and $\theta$. Replacing Eqs. (\ref{metricainversa}) and (\ref{raizdeg}) in Eq. (\ref{esq1q2}) gives
\begin{eqnarray} \label{eqsB0B1}
E \chi_{s}&=& \frac{1}{2m}\bigg[ -\frac{\hbar^2}{\rho(1+\rho^{'2})^{1/2}}\partial_{q_2}\left( \frac{\rho}{(1+\rho^{'2})^{1/2}}\partial_{q_2}\chi_{s} \right)+ \frac{iQ\hbar}{\rho(1+\rho^{'2})^{1/2}}\partial_{q_2}\left( \frac{\rho}{(1+\rho^{'2})^{1/2}}A_{q_2}\right)\chi_s \nonumber \\ &-&\frac{\hbar^2}{\rho(1+\rho^{'2})^{1/2}}\partial_{\theta}\left(\frac{(1+\rho^{'2})^{1/2}}{\rho}\partial_{\theta}\chi_{s} \right) + \frac{iQ\hbar}{\rho(1+\rho^{'2})^{1/2}}\partial_{\theta}\left(\frac{(1+\rho^{'2})^{1/2}}{\rho}A_{\theta}\right)\chi_s \nonumber \\ &+& \frac{2iQ\hbar A_{q_2}}{(1+\rho^{'2})}\partial_{q_2}\chi_s +\frac{Q^2 A^{2}_{q_2}\chi_s}{(1+\rho^{'2})}+\frac{2iQ\hbar A_{\theta}}{\rho^{2}}\partial_{\theta}\chi_s +\frac{Q^2 A^{2}_{\theta}\chi_s}{\rho^{2}} + V_{geo}\chi_s +QV\chi_{s}, 
\end{eqnarray}
where, for the sake of simplicity, we suppress the dependence of $\rho$ on $q_2$. Furthermore, we do not take into account the temporal character of Eq. (\ref{esq1q2}).

Eq. (\ref{eqsB0B1}) is the general equation for a charged particle confined to a cylindrical surface under the presence of magnetic and electric fields in any direction. This equation becomes quite complicated to solve for a magnetic field in any direction.

In Ref.\cite{ferrari2008cylindrical} this equation was solved for a magnetic field transverse to the axis of a straight cylinder (in this case, $\rho$ is constant). In our case,  we are interested in deformed nanotubes. Therefore,  $\rho$  can not be kept constant and this adds an extra  difficulty to the problem. We simplify the ``electromagnetic configutation'',  contemplating $V=0$ (absence of electric field) and a magnetic field along the cylinder axis, i.e. only in the direction of $q_2$. As a remarkable result,  Eq. (\ref{eqsB0B1}) becomes separable. This orientation of the magnetic field would not affect the dynamics of a charged particle if the radius of the nanotube were constant. However, in our case, the radius of the nanotube is varying, and as a consequence the charged particles are affected by the magnetic field as they pass through the deformations. 

With this choice, we have that $B_1=0$. So, the vector potential simplifies further
\begin{equation}
\vec{A}=\left(\frac{1}{2}\rho B_0 ,0,0 \right)
\end{equation}
and Eq. (\ref{eqsB0B1}) becomes
\begin{eqnarray} \label{eqsseparavel}
E\chi_s = \frac{1}{2m}\bigg[&-&\frac{\hbar^2}{(1+\rho^{'2})} \partial^{2}_{q_2}\chi_{s}-\frac{\hbar^2\rho^{'}}{\rho(1+ \rho^{'2})}\left( 1- \frac{\rho \rho^{''}}{(1+\rho^{'2})} \right)\partial_{q_2}\chi_{s}  \nonumber \\ &-&\frac{\hbar^2}{\rho^2}\partial^{2}_{\theta}\chi_{s} + iQ\hbar B_0 \partial_{\theta}\chi_{s}+\frac{Q^2 \rho^2 B_{0}^{2}}{4}\chi_{s} \bigg]+ V_{geo}\chi_{s}.
\end{eqnarray}
Eq. (\ref{eqsseparavel}) as mentioned above is separable. Therefore, we assume that the solution is of the type
\begin{equation}
\chi_{s}(q_{2},\theta)= \chi_{q_2}(q_{2})\chi(\theta),
\end{equation}
which consists of the product of two functions, one that depends only on $q_2$ and other that depends only on $\theta$. Using this ansatz, the angular part will be given by
\begin{equation} \label{parteangular}
\partial^{2}_{\theta}\chi(\theta) -il\partial_{\theta}\chi(\theta)+2l^2 \chi(\theta) = 0.
\end{equation}
The solutions of this equation are the eigenfunctions of the angular momentum $l\hbar$ along the axis $q_2$, and are of the type
\begin{equation}
 \chi(\theta)= e^{il\theta}. \label{circ}
\end{equation}
The axial part can be written as
\begin{equation} \label{equacaofinalmagnetica}
\partial_{q_2}^{2} \chi_{q_2} + F \partial_{q_2}\chi_{q_2} + G \left[ \frac{2m}{\hbar^2}(E-V_{geo})-\frac{l^2}{\rho^2} +\frac{QB_0 l}{\hbar}- \frac{Q^2 \rho^2 B_{0}^{2}}{4\hbar^2}\right] \chi_{q_2} = 0,
\end{equation} 
where
\begin{equation}
G\equiv 1+\rho^{'2} \hspace{1cm} \textrm{e} \hspace{1cm} F\equiv \frac{\rho^{'}}{\rho}\left[ 1 - \frac{\rho \rho^{''}}{(1+ \rho^{'2})} \right].
\end{equation}
Eq. (\ref{equacaofinalmagnetica}) is of the form	
\begin{equation} \label{forma}
\chi_{q_2}^{''} + V_1(q_2)\chi_{q_2}^{'}+V_2(q_2)\chi_{q_2} = 0, 
\end{equation}
where
\begin{equation}
V_1= F
\end{equation}
 and
\begin{equation}
V_2 = G \left[ \frac{2m}{\hbar^2}V_{geo} -\frac{l^2}{\rho^2} +\frac{QB_0 l}{\hbar}- \frac{Q^2 \rho^2 B_{0}^{2}}{4\hbar^2}\right].
\end{equation} 
Writing $\chi_{q_2}= \phi(q_2) \lambda(q_2)$, we obtain that
\begin{equation}
\phi^{\prime\prime} + \left(2\frac{\lambda^{\prime}}{\lambda} + V_1 \right) \phi^{\prime}  + \left(\frac{\lambda^{\prime\prime}}{\lambda}+ V_1 \frac{\lambda^{\prime}}{\lambda} +V_2 \right) \phi = 0.
\label{eqphi}
\end{equation}
Considering $2\frac{\lambda^{\prime}}{\lambda} + V_1 =0$, we get $\lambda(q_2) = e^{-\frac{1}{2}P(q_2)}$, where $P(q_2)$ is the primitive of $V_1$, which means that $P^{\prime}(q_2)=V_1(q_2)$. At this way, Eq. (\ref{eqphi}) becomes 
\begin{equation} \label{phimag}
\phi^{''}(q_2)+\left(- \frac{1}{4}V_{1}^{2}(q_2)-\frac{1}{2}V_{1}^{'}(q_2)+V_{2}(q_2) \right)\phi(q_2)=0.
\end{equation}
Here, we will call the term in parentheses an effective potential 
\begin{equation}
 V_{eff} = \left(- \frac{1}{4}V_{1}^{2}(q_2)-\frac{1}{2}V_{1}^{'}(q_2)+V_{2}(q_2) \right).
\end{equation} 

We consider the injection of electrons with energy $E_k$ from the negative part of the axis $q_2$. Thus, we have that the general solutions outside the deformation are plane waves given by
\begin{eqnarray} \label{ondasplanas}
\chi_{q_2} &=& a_0 e^{ik_0 q_2}+b_0e^{-ik_0 q_2},\hspace{2.5cm} q_2\leq 0 \nonumber \\ 
&=&  a_Le^{-ik_L(q_2 -L)}+b_Le^{ik_L(q_2 -L)}, \hspace{1 cm}  q_2\geq L,
\end{eqnarray} 
with $a_L=0$. We will consider that $V_{geo}(0) = V_{geo}(L)$, so
 \begin{equation}
 k_0 = \sqrt{\frac{2m}{\hbar^2}(E_k - V_{geo}(0))} = k_L.
\end{equation} 
Note that $V_1(q_2)=0$ for $q_2$ out of the range $0<q_2<L$, which makes Eq. (\ref{forma}) and Eq. (\ref{phimag}) identical. Then, Eq. (\ref{ondasplanas}) is also valid for $\phi(q_2)$, which means that $\chi_{q_2}(0)=\phi(0)$ and $\chi_{q_2}(L)= \phi(L)$. As a consequence, the reflection and transmission coefficients will not depend on $ \lambda(q_2)$ and can be obtained directly from $\phi(q_2)$.

With the boundary conditions $a_0=1$ and $a_L=0$, we have that
\begin{equation}
a_0 = \frac{1}{2}\left[ \phi(0) - i \phi^{\, '}(0)/k_0 \right]=1 \label{a0}
\end{equation}
and
\begin{equation}
a_L = \frac{1}{2}\left[ \phi(L) + i \phi^{\, '}(L)/k_L \right]=0 . \label{aL}
\end{equation}
Following the calculations done in Ref. \cite{santos}, we obtain that the transmission and reflection coefficients are given, respectively, by
\begin{equation} \label{transmag}
T=\frac{k_L}{k_0}|\phi (L)|^2
\end{equation}
and
\begin{equation}\label{refmag}
R=|\phi(0) -1|^2 . 
\end{equation}
Thus, the problem reduces to finding $\phi(0)$ and $\phi(L)$. This is done by solving the coupled differential and algebraic equations (\ref{eqphi}), (\ref{a0}) and (\ref{aL}) in the range $ 0 \leq q_2 \leq L $. {The fact that the magnetic field might be present outside this domain does not affect our results. There, we have an ordinary cylinder and the axial magnetic field effect is to add a phase to the plane wave solution, which is the same on both sides of the deformation. Nevertheless, we point out that this does not apply in the case of a varying magnetic field \cite{onorato2013carbon}.}

With the above equations, we implemented a code in MAPLE to find, for each injected energy, $ \phi(0) $ and $ \phi(L) $ and, consequently, the transmission and reflection coefficients \cite{santos,piropo2020surfing}. We use a mixed unit system in which we have the incident energy in meV, and the distances in nm. At this way, the electron mass is given by $m_e = 5.68 \times 10^{-27}$ meV$\cdot$s$^2$/nm$^2$ and the Planck constant $\hbar = 6.58 \times 10^{-13}$ meV$\cdot$s. Also, for the magnetic field, we have the unit meV$\cdot$s/C$\cdot$nm$^2$. 

In the next section, we will present our results, where we investigate the influence of a magnetic field in the transport properties of various deformed nanotubes, where we will make a comparison between the cases with and without magnetic field.

\section{Numerical Results and Discussions}

In this section, we analyze the effects of the presence of the magnetic field in the transport properties of deformed nanotubes. We consider that the deformation in the nanotubes are corrugations generated by the curve
\begin{equation}
\rho(q_2)= R + \frac{R \epsilon}{2} \left[1 -\cos \left(2 \frac{n \pi q_2}{L} \right) \right],
\end{equation} 
where $R$ is the initial and final radius of the nanotube, $\epsilon$ gives the strength of the increase (positive values) or decrease (negative values) of the radius of the nanotube, $L$ is the length of the corrugated region, and $n$ is the number of corrugations. It is important to mention that such corrugations were already investigated in a plane \cite{PhysRevB.79.235407} and nanotubes \cite{santos,SERAFIM2019139}, but the analysis of the influence of a magnetic field has not been addressed yet.

 {The theoretical approach considered here describes a free particle constrained to a cylindrical surface. It can be used, for instance, to describe the electronic transport in a metallic cylinder. {The most direct application of the model is in C2DEGs. However, it can also be applied, for instance, for metallic WS$_2$ and carbon nanotubes.} Due to the low-energy linear dispersion of metallic CNT, they are well described by an effective Dirac equation. However, in the ballistic limit, the wavefunction for the scattering by a one-dimensional potential can be obtained from the Schr\"odinger equation \cite{saito1998physical}. So, our approach can also be used in such systems. In what follows, we will consider} $R=75$~nm, {which is a radius that can be realized in C2DEGs and also in CNTs, as in the CNT NT145 obtained in Ref. \cite{NagaiE1330}}. {It is important to mention that, in contrast to graphene, the spin-orbit coupling plays an important role in the transport properties of CNTs \cite{RevModPhys.87.703}. However, the magnitude of the spin-orbit coupling is inversely proportional to the diameter of the CNT. Therefore, since we are considering a CNT with a very large diameter, the spin-orbit interaction can be neglected.} For the sake of clarity, we will show in the results the magnetic field in tesla, where $1$~T$=6.25 \times 10^3$~meV$\cdot$s/C$\cdot$nm$^2$. We will also consider that the charge carriers are electrons, which means that $Q=-e$.


\begin{figure*}[h!]
\centering
\includegraphics[angle=0,width=5 cm,height=4.5 cm ]{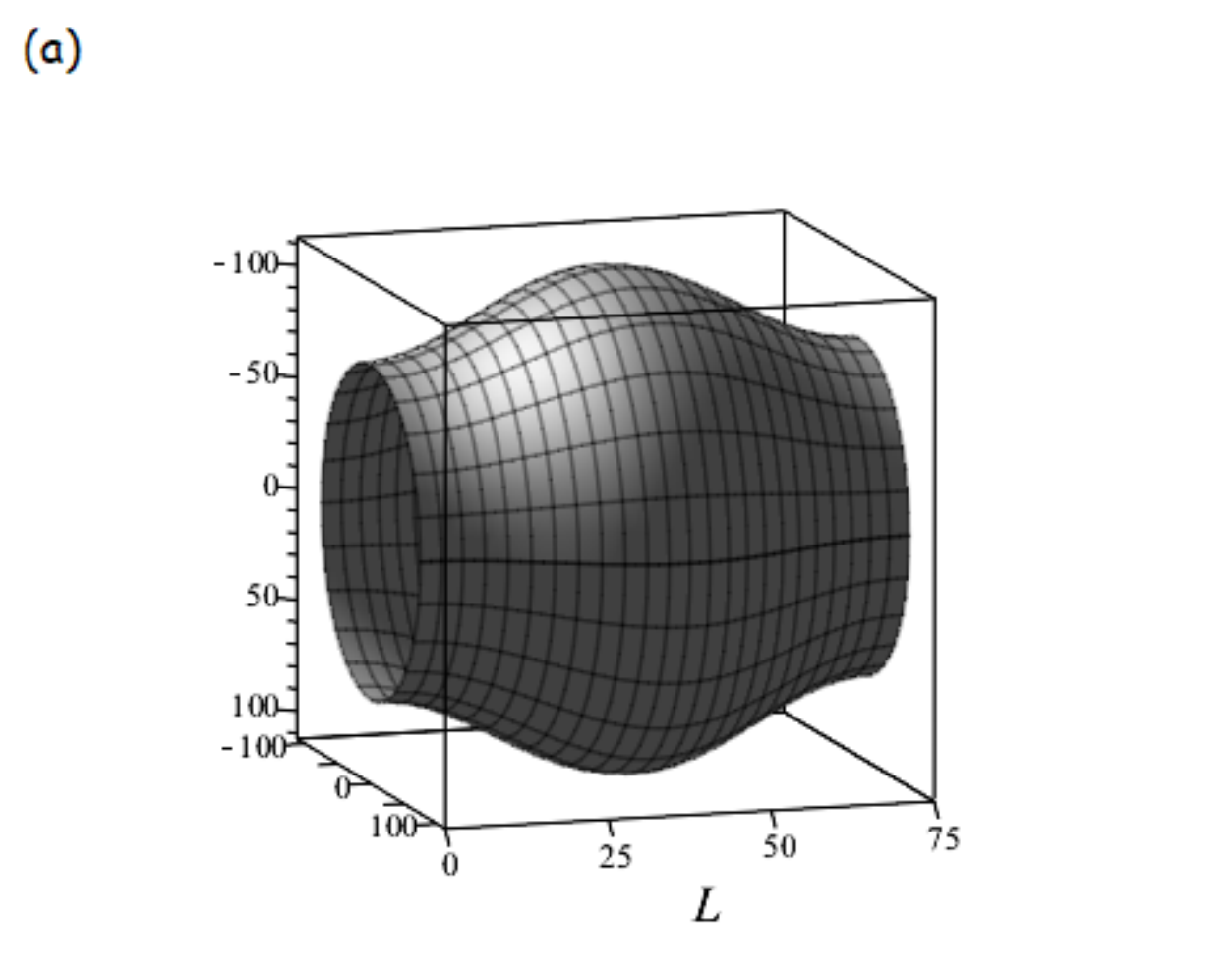}
\includegraphics[angle=0,width=5.6 cm,height=4.5 cm ]{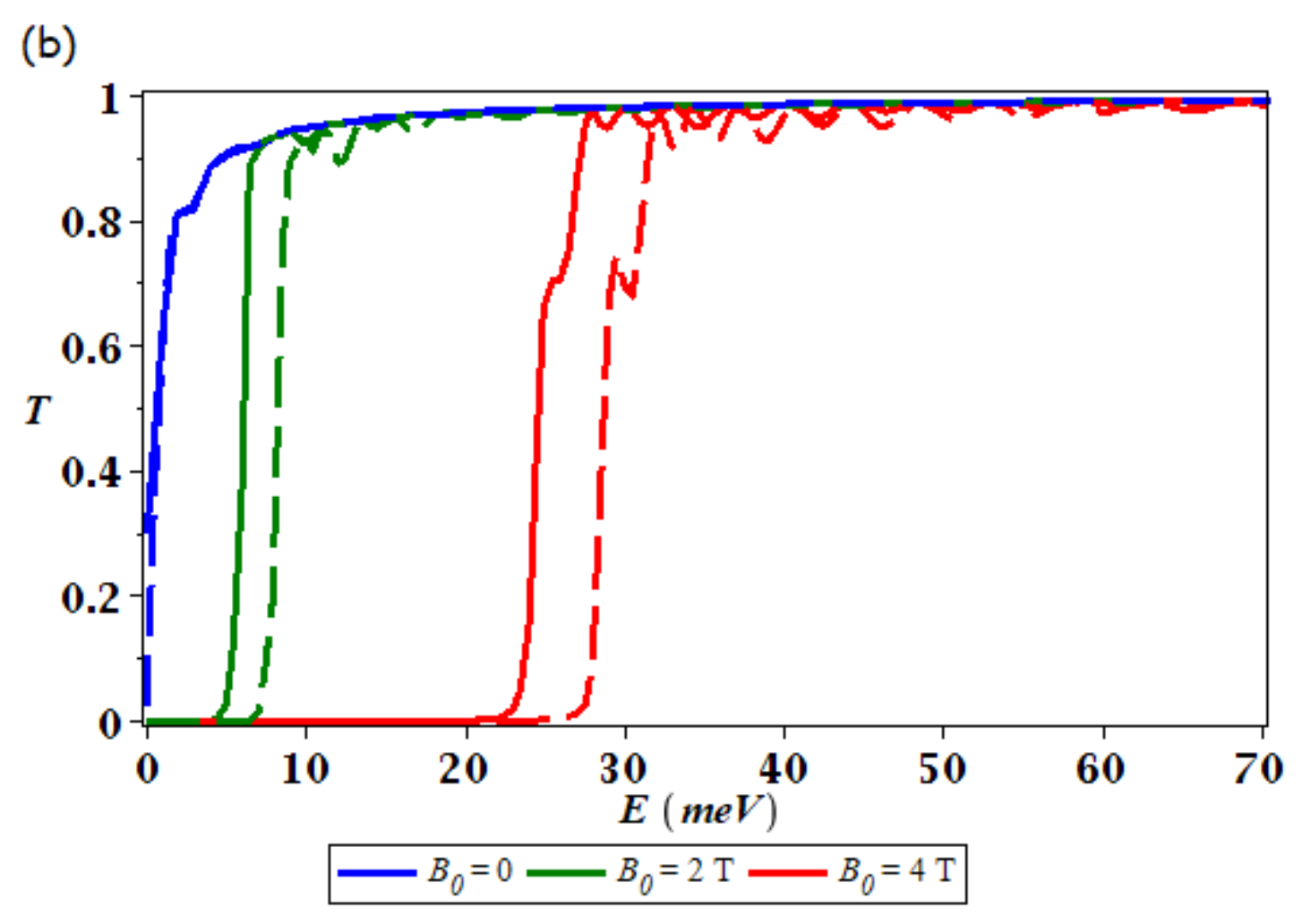}
\includegraphics[angle=0,width=5.6 cm,height=4.5 cm ]{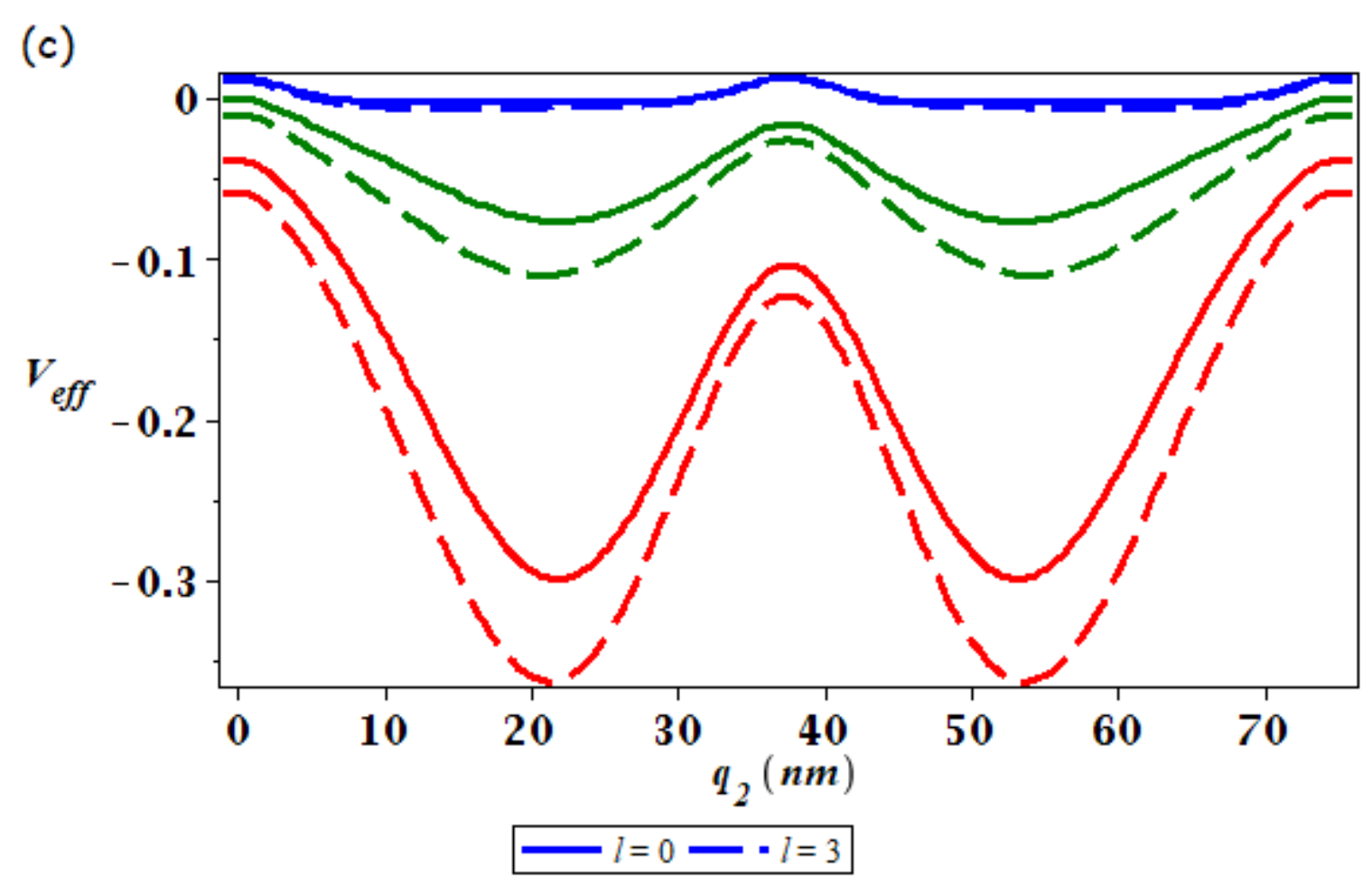}
\caption{(b) The transmittance and (c) the effective potential for one bump with $\epsilon=0.5$. We consider here $B_0=0$ (blue lines), $B_0=2$~T (green lines) and $B_0=4$~T (red lines). We also consider $l=0$ (continuous lines) and $l=3$ (dashed lines).}
\label{1bump}
\end{figure*}

\begin{figure*}[h!]
\centering
\includegraphics[angle=0,width=5 cm,height=4.5 cm ]{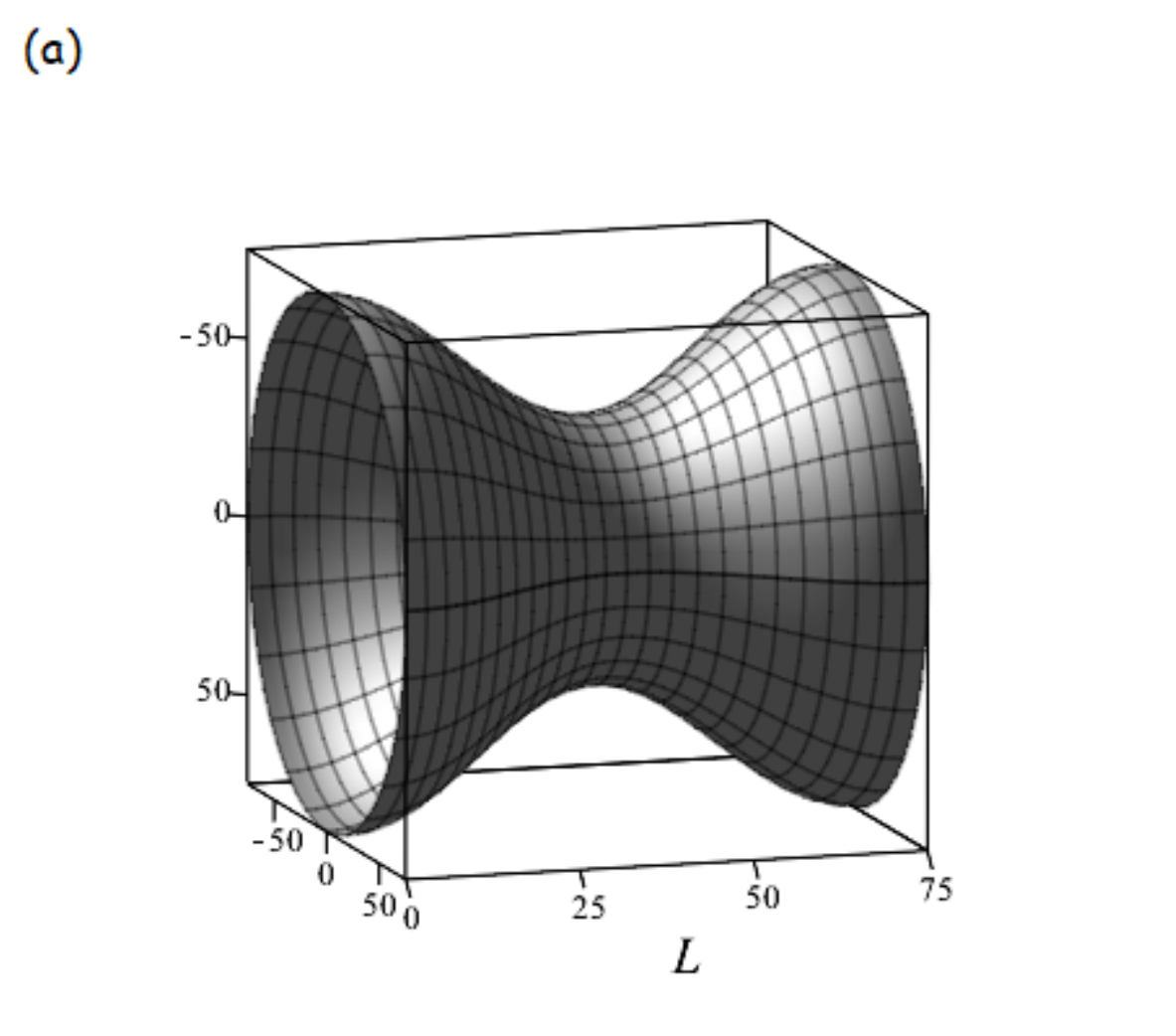}
\includegraphics[angle=0,width=5.6 cm,height=4.5 cm ]{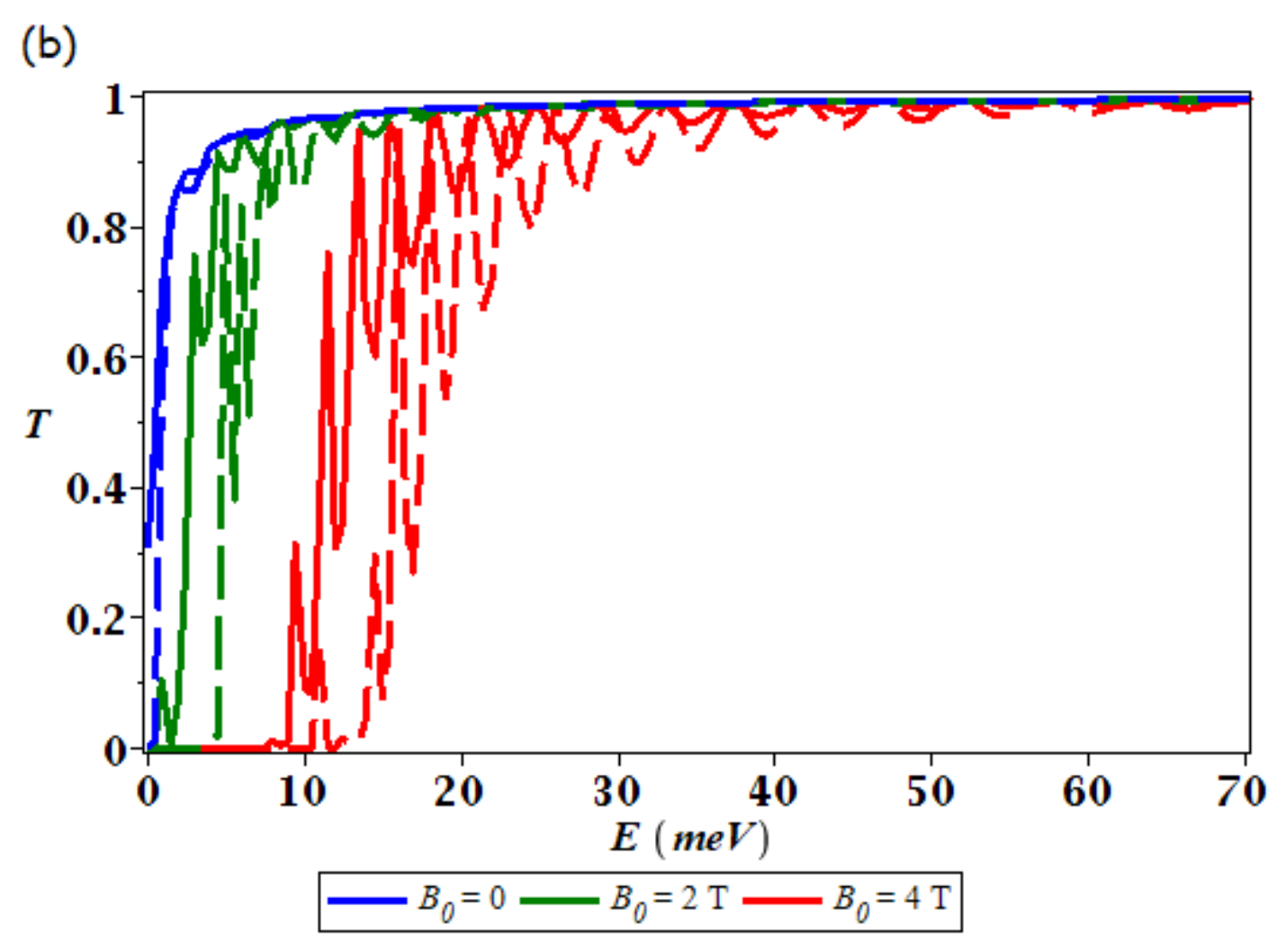}
\includegraphics[angle=0,width=5.6 cm,height=4.5 cm ]{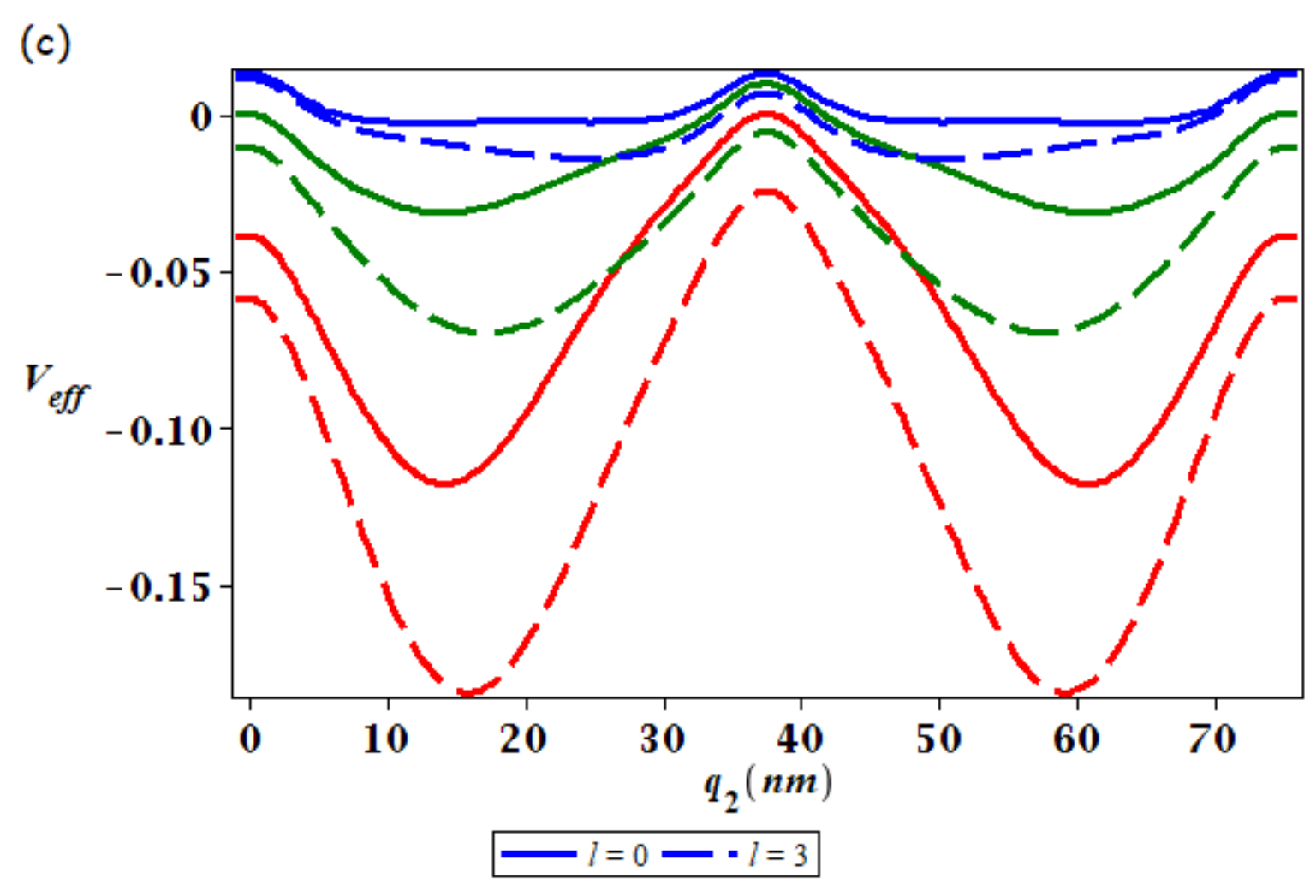}
\caption{(b) The transmittance and (c) the effective potential for one depression with $\epsilon=-0.5$. We consider here $B_0=0$ (blue lines), $B_0=2$~T (green lines) and $B_0=4$~T (red lines). We also consider $l=0$ (continuous lines) and $l=3$ (dashed lines).}
\label{1dep}
\end{figure*}

In Figs. \ref{1bump} and \ref{1dep}, we analyze the influence of the magnetic field in the transmittance of a nanotube with one bump and one depression, respectively. The parts $(b)$ and $(c)$ of the figures show the transmittance and the effective potential, respectively. {Note the similarity between the effective potential of  the bump and the one of the depression, even though they have opposite Gaussian curvatures. This is due to the fact that $V_{geo}$ (see Eq. \eqref{Vgeo} ) is always attractive.} We consider three different values for the magnetic field ($B_0=0$~T, $2$~T, $4$~T) and two values for the angular momentum ($l=0,3$). {We note that the typical magnetic length scale $(\hbar/eB)^{1/2}$ is of the order of $10$~nm for the values of the magnetic field considered here, therefore of the typical scale of the corrugation length $\simeq 30$~nm or of the cylinder radius, $75$~nm.} We can clearly note that the transmittance is sensitive to the magnetic field. In fact, the transmittance curve is shifted to the right with the introduction of the magnetic field: {the device behaves as a high-pass filter}. This displacement is because the effective potential is now deeper, causing the less energetic electrons to be reflected or trapped in the potential. The same occurs when we increase the angular momentum. However, the shift in the transmission observed when $l$ increases is greater as the $B_0$ increases. The deeper quantum wells seen in the case of bumps is a consequence of the fact that the term in the effective potential that contains the geometrical potential becomes positive for the case of depressions, reducing the depth of the wells.

Note also the appearance of resonance peaks, which correspond to quasi-bound states. These states are associated with a quantum well where a particle is primarily confined, but has a finite probability of tunnelling and escaping. In the nanotube, the effective potential (containing the geometric potential), if deep enough, there may be similar states. In the cases where the energy of the charged carriers coincides with that of a quasi-bound state, it easily tunnels into the potential region, and so tunnel to the opposite side. Even though the bumps induce deeper wells, a higher oscillation in the transmission is observed for the depressions. It occurs because a potential barrier appears in the effective potential in the middle of the depression ($q_2=37.5$~nm in our case). {We remark that the resonance peaks in the transmittance are not  solely due to the magnetic field. They also appear  in the absence of the field, due to the geometric potential,  as can be seen in Figs. 1(c) and 2(c) of Ref. \cite{santos}.  }


\begin{figure*}[h!]
\centering
\includegraphics[angle=0,width=5.4 cm,height=5.4 cm ]{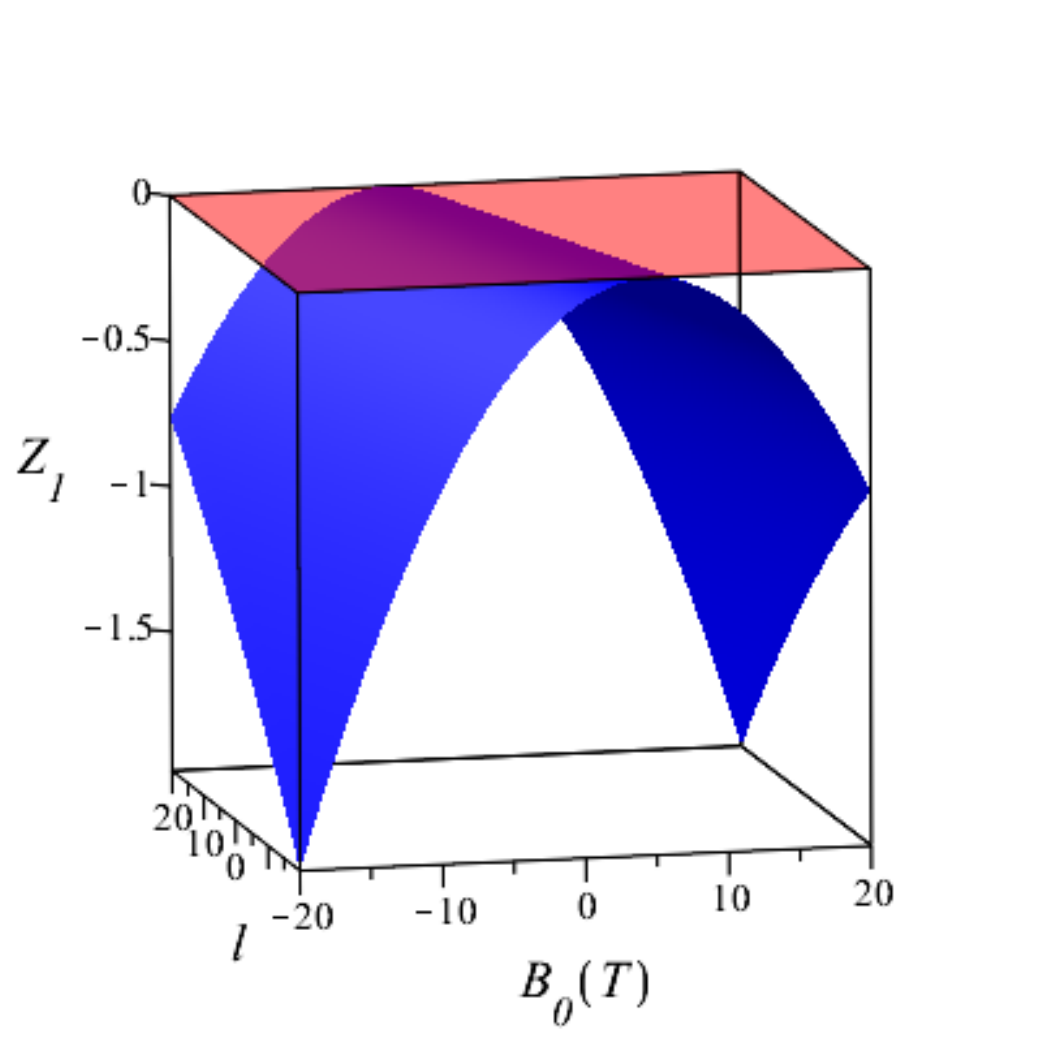}
\includegraphics[angle=0,width=5.4 cm,height=5.4 cm ]{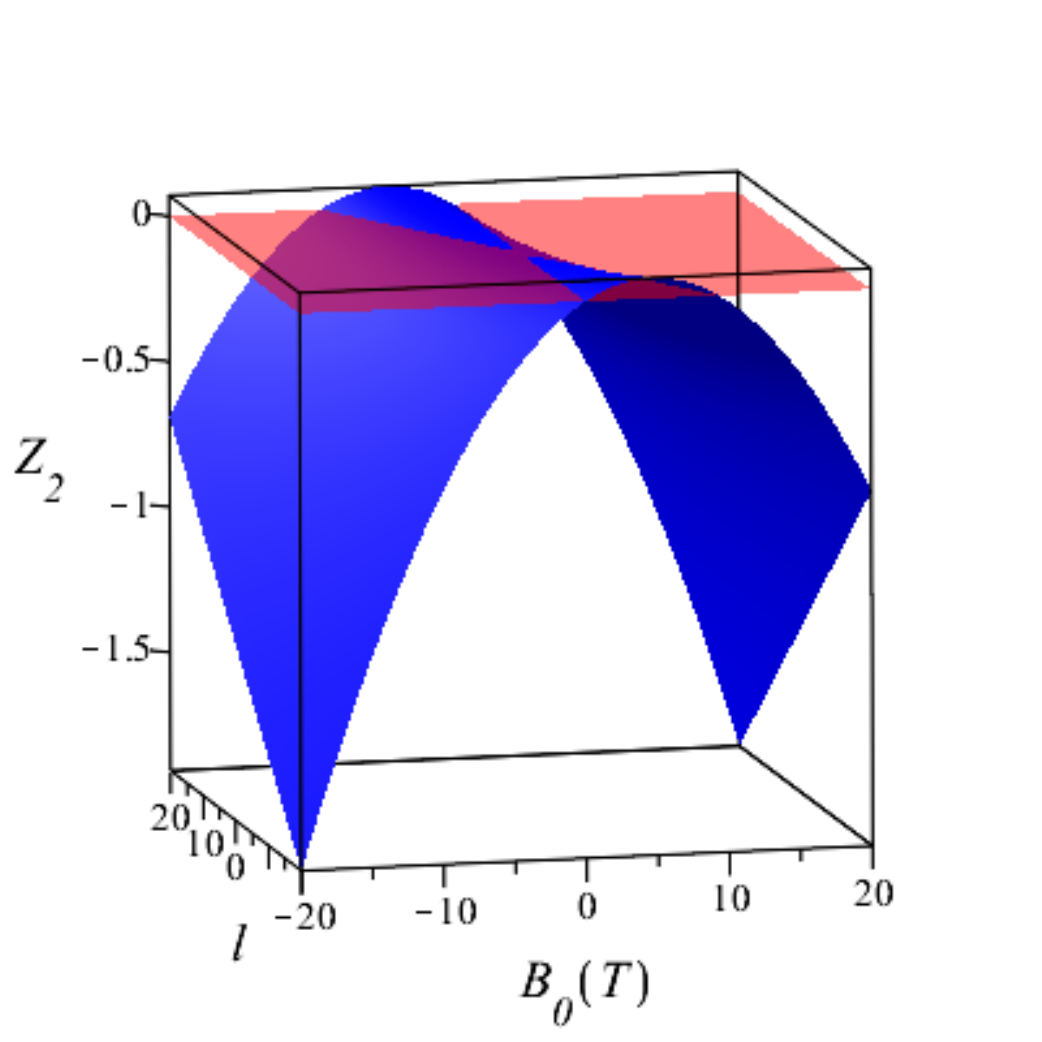}
\includegraphics[angle=0,width=5.4 cm,height=5.4 cm ]{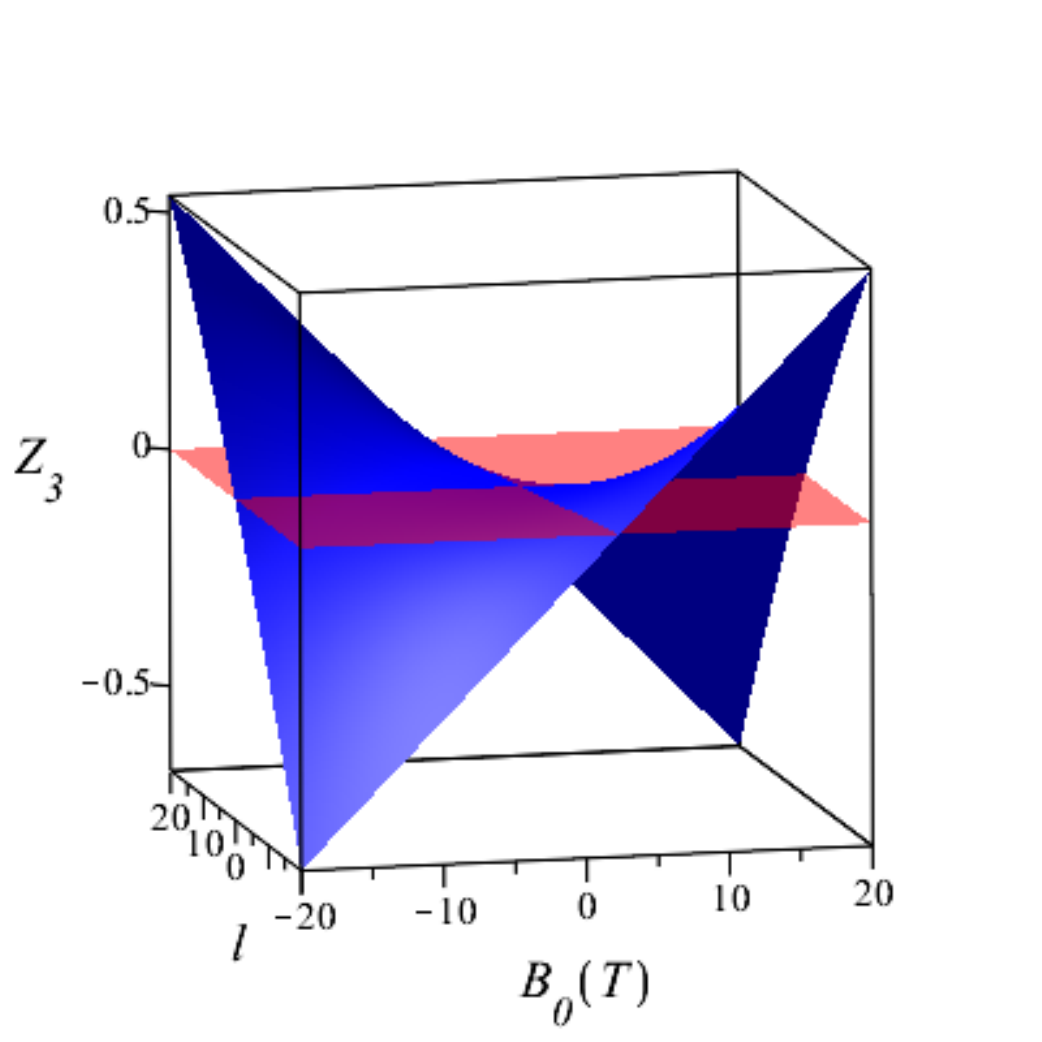}
\caption{From left to right: $Z_1$, $Z_2$ and $Z_3$ as a function of $l$ and $B_0$ (blue surfaces) with $\rho=75$~nm. The red planes reveal when $Z_1$, $Z_2$ and $Z_3$ are equal to zero.}
\label{3d}
\end{figure*}

In order to clarify the influence of the magnetic field in the transport properties of the deformed carbon nanotubes, let us analyze the three terms in the effective potential that depend on $l$ and/or $B_0$. These terms are shown in the equation below:
\begin{equation}
 Z_1= -\frac{l^2}{\rho^2} -\frac{eB_0 l}{\hbar}- \frac{e^2 \rho^2 B_{0}^{2}}{4\hbar^2}.
\end{equation}
In Fig. \ref{3d} $Z_1$ is plotted as a function of $l$ and $B_0$. We can see that the angular momentum and the magnetic field can only induce a deeper effective potential, since $Z_1$ can not have a positive value. Also, for each value of $l$ there is a value of $B_0$ that $Z_1=0$. It means that the magnetic field can be used to choose which angular momentum will cross more easily the deformations, since a higher transmission is obtained for $Z_1=0$. In other words, there is a value of $B_0$ that makes an electron with a specific angular momentum cross the deformations as if it had no angular momentum and there is no magnetic field applied. It occurs when $B_0=-2\hbar l/(e\rho^2)$. It is very important to remember here that $\rho$ depends on the position. So, this last relation has to be used carefully. As we will see below, there is a constant value of $\rho$ between its maximum and minimum values in the deformed region that satisfies this relation. Then, in fact, the value of $\rho$ that can be used in this relation depends on the deformation of the surface. 

We also define
\begin{equation}
 Z_2=-\frac{eB_0 l}{\hbar}- \frac{e^2 \rho^2 B_{0}^{2}}{4\hbar^2},
\end{equation}
which consider the two terms in the effective potential that depend on the magnetic field. $Z_2$ as a function of $l$ and $B_0$ can be seen in Fig. \ref{3d}. When $Z_2=0$, the charge carriers do not see the magnetic field. So, it is possible to choose a specific angular momentum that will not feel the influence of the magnetic field. It occurs for a given $l$ when $B_0=-4\hbar l/(e\rho^2)$. Again, remember that $\rho$ depends on the position.

Finally, we define 
\begin{equation}
 Z_3= -\frac{l^2}{\rho^2} -\frac{eB_0 l}{\hbar},
\end{equation}
which is the combination of the two terms in the effective potential that depend on the angular momentum. The plot of $Z_3$ in terms of $l$ and $B_0$ is also shown in Fig. \ref{3d}. As can seen, for a given value of $l$, there is a magnetic field that cancels the influence of the angular momentum in the effective potential. It happens when $B_0=-\hbar l/(e\rho^2)$.

\begin{figure}[!htb] 
\centering
\includegraphics[width=\linewidth]{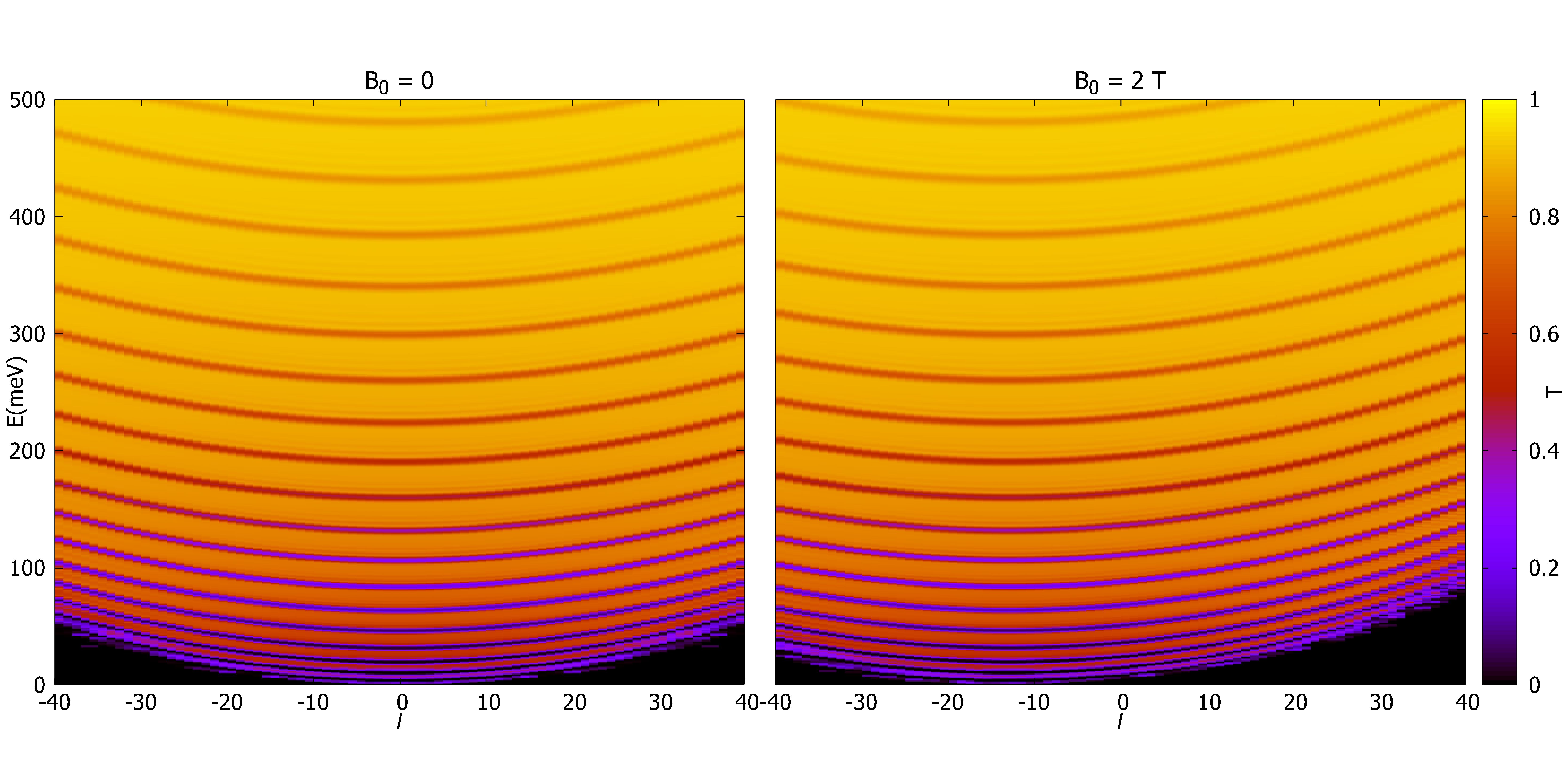}
\caption{Contour plot of the transmittance as a function of $l$ and $E$ with $B_0=0$ and $B_0=2$~T for 3 bumps with $\epsilon=0.5$.}
\label{contourbump}
\end{figure}

\begin{figure}[!htb] 
\centering
\includegraphics[width=\linewidth]{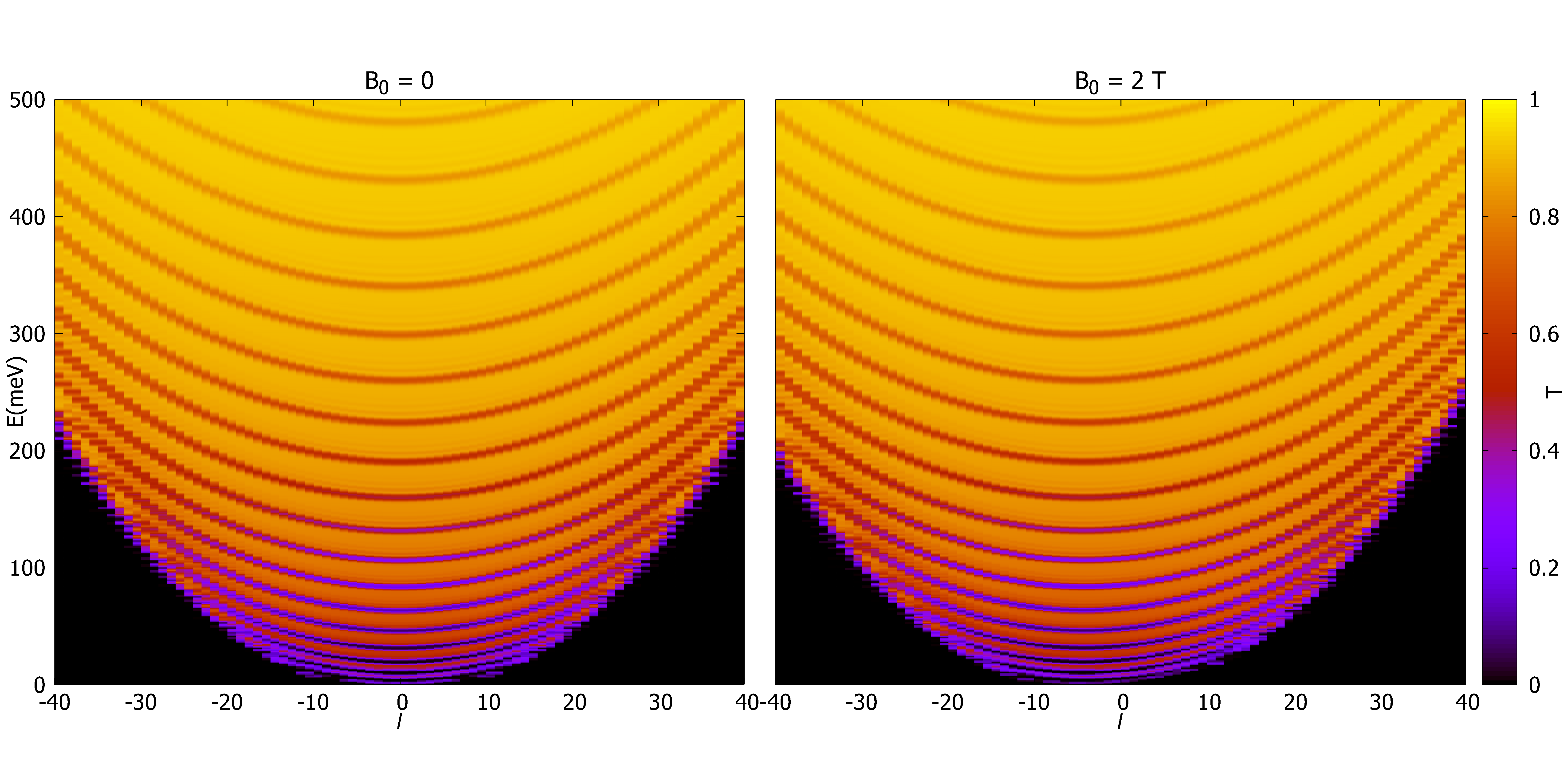}
\caption{Contour plot of the transmittance as a function of $l$ and $E$ with $B_0=0$ and $B_0=2$~T for 3 depressions with $\epsilon=-0.5$.}
\label{contourdep}
\end{figure}

The influence of the angular momentum and the magnetic field in the transmission that were discussed here can be seen more clearly in Figs. \ref{contourbump} and \ref{contourdep}, where we have the transmittance as a function of the energy and the magnetic field for different values of $B_0$ for the cases of three bumps and three depressions, respectively. We can see a parabolic contour of the transmittance. For $B_0=0$ the vertex of the parabola is at $l=0$, which is the angular momentum that crosses the deformation more easily. When $l$ changes, the transmittance is shifted up, opening regions with no transmission for low energies. When $B_0$ increases, the whole contour is shifted to the left. With a negative value of $B_0$, which means a magnetic field in the negative direction of the axis $q_2$, this shift would occur to the right side. 

Comparing the cases of bumps and depressions, we can see that each one has its advantages. The shift induced in the transmittance when $l$ changes is more significant for the case of depressions. It reveals that deformations with negative values of $\epsilon$ are more suitable to filter low energy charge carriers. However, the magnetic field can be used to select which angular momentum will cross the deformations more easily. And the deformations with positive values of $\epsilon$ are more sensitive to a change of $B_0$. In fact, the shift induced by a change in the magnetic field $\Delta B_0$ is given by $\Delta l= -e\rho^2 \Delta B_0 /(2\hbar)$. Since the radius $\rho$ of the nanotubes is reduced in case of depressions, more significant shifts of the transmittance contour are expected in comparison with the case of bumps.

\section{Conclusions}

In this work, we investigated some properties of electronic transport on a device consisting of a corrugated  {metallic} nanotube submitted to an external magnetic field. The magnetic field is oriented along the cylinder axis for simplicity. The corresponding Schrödinger equation was built for a spinless charge, confined to an axisymmetric shell through an effective potential (da Costa procedure). {With or without the magnetic field, the solution is separable. The motion on the circumference is given by Eq. \eqref{circ} and the motion along the axis by the solutions of Eq.\eqref{phimag}, which we have found numerically.} Our model showed that the magnetic field and the curvature could be adequately encompassed in a global effective potential to which electrons strongly couple. 

Numerical simulations were run to compute the longitudinal transmittance of the nanotube in the presence of bumps and depressions. They reveal that the device acts as a high-pass filter which inhibits the flow of low-energy electrons. Moreover, the effective potential also favours the transmission of electrons endowed with a tunable value of their orbital angular momentum $l$. An additional coupling between the magnetic field and the electron angular momentum was also mentioned, which enhance transmittance levels when $l$ is negative.    

In summary, $B$ can thus be used to tune the properties (energy, orbital angular momentum) of the transmitted electron flux. Electrons also posses a spin degree of freedom, and it is now recognized that spin-orbit interactions can be significantly high on curved carbon nanotubes \cite{Steele2012}. Hence, the possibility to tune spin from the external magnetic field makes our device promising for
many applications, such as the generation of carbon nanotube spin qubits. This will be the subject of further investigations. {A further development of this work could be the application of an electric field to the system. The field $\vec{E} =-\vec{\nabla}V $ contributes to the Schr\"odinger equation  \eqref{eqsseparavel} with a term $QV\chi_{s}(q_{2},\theta)$ (see Eq. (13) of Ref. \cite{Ferrari}). In the simplest case, where the field is parallel to the tube, Eq. \eqref{eqsseparavel} remains separable and the electric field contribution appears only in the motion along $q_2$. That is, $QV(q_{2})$ will contribute to $V_2$ in Eq. \eqref{phimag}, changing thus the effective potential $V_{eff}$. Depending on the functional form of $V(q_2)$ new minima may appear in the effective potential leading to new transmittance peaks, since the peaks are associated to quasi-bound states. More likely, the already present minima due to the geometric and magnetic contributions will be shifted by the introduction of the electric potential, which will also shift the transmittance peaks, thus directly affecting the conductance.}

{A more subtle development of the model used here, the study of the Aharonov-Bohm-related effects in a deformed nanotube, is certainly worth pursuing. To do this, the magnetic field must be restricted to the interior of the tube (a threading Aharonov-Bohm flux) as studied in \cite{onorato2013carbon}. As described in this reference, for straight nanotubes, there appear oscillations in the longitudinal ballistic and persistent currents. It would be interesting to study the consequences on these properties of the geometric potential introduced by a deformation in the tube.}

\begin{acknowledgements}
F.M. thanks CNPq. JRFL thanks Capes, CNPq and Alexander von Humboldt Foundation. 
\end{acknowledgements}

\subsection*{Data Availability Statement}
Data sharing is not applicable to this article as no new data were created or analyzed in this study.

\bibliography{ref}

\end{document}